\documentclass[12pt,letterpaper]{JHEP3}

\usepackage{amssymb}

\def\be{\begin{equation}}
\def\ee{\end{equation}}
\def\barr{\begin{array}{lr}}
\def\earr{\end{array}}
\def\bea{\begin{eqnarray}}
\def\eea{\end{eqnarray}}

\def\del{\partial}

\def\o{{\cal O}}

\def\S{{\cal S}}

\def\sl2{$SL_2({\mathbb R})$}

\def\pa{\partial}
\def\del{\nabla}

\def\to{\rightarrow}
\def\be{\begin{equation}}
\def\ee{\end{equation}}
\def\bea{\begin{eqnarray}}
\def\eea{\end{eqnarray}}
\def\nonu{\nonumber \\{}}
\def\half{{1 \over 2}}

\def\sF{{{\rm F}\!\!\!\!\hskip.8pt\hbox{\raise1pt\hbox{/}}\,}}

\def\a{\alpha}

\def\d{\delta}
\def\e{\epsilon}
\def\ve{\varepsilon}
\def\f{\phi}
\def\g{\gamma}
\def\h{\eta}

\def\k{\kappa}

\def\m{\mu}
\def\n{\nu}
\def\o{\omega}
\def\p{\pi}
\def\q{\theta}

\def\s{\sigma}
\def\t{\tau}

\def\x{\xi}

\def\F{\Phi}
\def\G{\Gamma}

\def\O{\Omega}

\def\S{\Sigma}

\title{Supersymmetric D-branes in the D1-D5 background}

\author{\centerline{Joris Raeymaekers$^1$ and K.P. Yogendran$^{2,3}$}\\

\centerline{$^1$: Department of Physics, University of Tokyo,}
\centerline{Hongo 7-3-1, Bunkyo-ku, Tokyo 113-0033, Japan.}

\centerline{$^2$: Center for Quantum Spacetime (CQueST),}
\centerline{Sogang University, Mapo-gu, Seoul, 121-742, Korea}
\centerline{$^3$: Dept. of Physics, University of Seoul,}
\centerline{Dongdaemun-gu, Seoul, 130-743, Korea.}
\bigskip
\centerline{{\rm E-mail}:\email{joris@hep-th.phys.s.u-tokyo.ac.jp}}
\centerline{{\rm E-mail}:\email{kpy@theory.tifr.res.in, pattag@gmail.com}}}

\abstract{We construct supersymmetric D-brane probe solutions in the background of the 2-charge D1-D5 system
 on $M$, where $M$ is either $K3$ or $T^4$. We focus on `near-horizon bound states' that preserve
supersymmetries of the near-horizon $AdS_3 \times S^3 \times M$ geometry and are static with respect to the global
time coordinate. We find a variety of half-BPS solutions that span an $AdS_2$ subspace in $AdS_3$, carry worldvolume flux
and can wrap an $S^2$ within $S^3$ and/or supersymmetric cycles in $M$.
}


\preprint{hep-th/0607150\\UT-06-14}

\begin{document}

\section{Introduction and summary}

The study of D-brane probes in the near-horizon region
of BPS D-brane systems has recently had interesting applications to the counting of BPS degeneracies.
D-brane systems with an $AdS_p \times S^q$ near-horizon region where supersymmetry is enhanced
allow for D-brane probe configurations localized near the horizon preserving a portion of the
enhanced supersymmetries.
Such branes have been  constructed  in backgrounds with $AdS_2 \times S^2$ \cite{Simons:2004nm},  $AdS_3 \times S^2$
\cite{Gaiotto:2006ns}
and $AdS_2 \times S^3$ \cite{Li:2006uq} near-horizon geometries. The branes considered in these papers possess a
number of interesting
properties. The solutions of interest are static with respect to
a choice of global time coordinate, and supersymmetry fixes their radial position in $AdS$ in terms of their charges.
Furthermore, they preserve
half of the near-horizon supersymmetries but break all of the supersymmetries of
the full asymptotically flat geometry. In case they wrap the sphere or a cycle in
the internal space with worldvolume flux turned on, they carry lower D-brane charge and can be
seen as bound states of lower-dimensional D-branes through a form of the Myers effect \cite{Gaiotto:2004pc,Rodriguez-Gomez:2005na}.

It is natural to interpret such branes as `near-horizon bound states' of the D-brane system, and one would expect
that by quantizing their moduli one should be able to count degeneracies of BPS states.
This expectation was borne out for the D0-D4 black hole in type
IIA, where the quantum mechanical counting reduces to counting lowest Landau levels in a magnetic field on the internal
space and reproduces
the  entropy both for the `large' \cite{Gaiotto:2004ij} and `small' \cite{Kim:2005yb} black hole
cases. Furthermore,
for black holes constructed out of M5-branes, the elliptic genus can be reconstructed by counting
near-horizon wrapped membrane states \cite{Gaiotto:2006ns,Gaiotto:2006wm}.

Motivated by these results, we revisit the two-charge D1-D5 system on $M$ (where $M$ can be $T^4$ or $K_3$), forming
a black string in 6 dimensions (see \cite{Aharony:1999ti,Justin} for reviews).
This system has a large ground-state degeneracy, the logarithm of which is proportional to $\sqrt{Q_1 Q_5}$ for
large charges.
We will look for supersymmetric D-branes in the near horizon $AdS_3\times S^3 \times M$ region.
In earlier works, half-BPS solutions which carry momentum along certain directions, such
as giant gravitons \cite{GG} and branes wrapping $S^3$ with momentum along $AdS_3$ \cite{Das} were
constructed. The branes we will consider here differ from these in that they do not carry any
momentum and are entirely static with respect to global time. We allow the branes to carry arbitrary worldvolume
fluxes (and hence also induced lower-dimensional D-brane charges).

The 16 supersymmetries of the near-horizon region split into 8 supersymmetries that extend
to the full asymptotically flat solution, which we will call `Poincar\'{e} supersymmetries',
and 8 `enhanced' supersymmetries that exist only in the near-horizon limit.
They are most easily distinguished in Poincar\'{e} coordinates in $AdS_3$.
D-brane probes
preserving some Poincar\'{e} supersymmetries should have  a BPS counterpart in the full geometry,
 and we will verify
for each solution whether it preserves Poincar\'{e} supersymmetries.

The outcome of our classification yields a large variety of D-branes preserving half of the
near-horizon supersymmetries  and is summarized in the following table.
\begin{center}
\begin{tabular}{|c|c|c|c|c|c|}\hline
brane type & $AdS_3$ & $S^3$ & $M$ & near-horizon susy & Poincar\'{e} susy   \\ \hline \hline
D1 & $AdS_2$ & $\cdot$ & $\cdot$ & 1/2 & 1/2 \\ \hline
D3 & $AdS_2$ & $\cdot$ & 2-cycle & 1/2 & 1/2 \\ \hline
D5 & $AdS_2$ & $\cdot$ & $M$ & 1/2 & 1/2 \\ \hline \hline
D3 & $AdS_2$ & $S^2$ & $\cdot$ & 1/2 & 1/2 \\ \hline
D7 & $AdS_2$ & $S^2$ & $M$ & 1/2 & 1/2 \\ \hline \hline
\end{tabular}
\end{center}
The solutions come in two types: branes of the first type span an $AdS_2$ subspace in $AdS_3 \times S^3 $ (and
possibly wrap a supersymmetric cycle in
$M$) while the second type of branes spans an $AdS_2 \times S^2$ subspace in $AdS_3 \times S^3$ (and possibly
wrap the whole of $M$). Branes of the second type are dipolar as the $S^2$ is contractible within $S^3$, and are
stabilized by worldvolume
flux \cite{Bachas:2000ik,Pawelczyk:2000hy}. The size of the $S^2$ is quantized in terms  of the number of fundamental strings bound to
the D-brane.
In all the above  solutions, the radial position in $AdS_3$ is fixed in terms of the charges and it is natural to
view them as
`near-horizon bound states' of the D1-D5 system. One novel feature is that, contrary to the examples in other backgrounds
discussed above,
these probe branes do preserve half  of the Poincar\'{e}
supersymmetries of
the full asymptotically flat geometry.

Let us comment on  related D-brane solutions that have appeared in the literature. The $AdS_2 \times S^2$ branes
were studied from the point of view of the DBI action in \cite{Pawelczyk:2000hy,Couchoud:2003jw}.
There is a substantial body of work discussing D-branes in the S-dual F1-NS5 background starting with \cite{Bachas}.
The S-dual versions of branes with worldvolumes
$AdS_2$ and $AdS_2 \times S^2$ appear there (the latter was shown to be half-BPS). A sampling of
further studies of $AdS_2$ branes in the
NS background includes \cite{Bachas:2002nz}
We also want to point out that D-branes with an $AdS_2$ component to their worldvolume are known to exist in other D-brane
backgrounds as well \cite{Camino:2001at,Skenderis:2002vf,Drukker:2005kx,Gomis:2006sb}.


This paper is organized as follows. In section \ref{background} we construct the Killing spinors on
$AdS_3\times S^3 \times M$ in suitable coordinates and review the conditions for probe branes to preserve supersymmetry.
In section \ref{ads2branes} we construct supersymmetric branes which extend along $AdS_2$ and possibly wrap cycles on $M$.
In section \ref{ads2s2branes} we turn to branes that span $AdS_2 \times S^2$ and possibly wrap cycles on $M$.
We end with a discussion of open problems in \ref{discussion}.
Appendix \ref{appendix} discusses how  Poincar\'{e} supersymmetries extend to the full geometry and
appendix \ref{otherbranes} discusses some branes spanning other submanifolds, none of which were found to be supersymmetric.

\section{The near-horizon limit of the D1-D5 system} \label{background}
In this section, we review some properties of the near-horizon limit of the D1-D5 system
needed in the rest of the paper.
\subsection{Background}
Consider type IIB on $S^1 \times M$ (with $M$ being either $K_3$ or
$T^4$), with D5-branes wrapped on $S^1 \times M$ and D1-branes on
$S^1$. We will take the $S^1$ radius  to infinity in what follows, so that the configuration
looks like a black string in six dimensions. The near-horizon supergravity background is $AdS_3 \times S^3 \times M$
with constant dilaton and nonvanishing RR 3-form flux $F^{(3)} = d C^{(2)}$.
On $AdS_3$, we will use a global `anti-de
Sitter' coordinate system $(\t, \o, \x )$ in which the constant $\x$ slices are
isomorphic to $AdS_2$. The supergravity background is then given by (see e.g. \cite{Maldacena:1998bw}):
\bea
ds^2 &=& r_1 r_5 [ d \x^2 + \cosh^2 \x (- \cosh^2 \o d\t^2 + d\o^2)
\nonu &&
+ d \psi^2 + \sin^2 \psi ( d \theta^2 + \sin^2 \theta d\varphi^2) ] +
{r_1 \over r_5} ds^2_{M}\nonu
e^{-\f} &=& {1 \over g} {r_5 \over  r_1}\nonu
C^{(2)} &=& {r_5^2 \over g} [ (\x + \half \sinh 2 \x ) \cosh \o d\o \wedge d\t +
(\psi - \half \sin 2 \psi) \sin \theta d\theta \wedge d\varphi ] \label{globalcoords}
\eea
where $ds^2_{M}$ is a Ricci-flat metric on $M$ and
\bea
r_5 &=& \sqrt{ g Q_5 \a '} \nonu
r_1 &=& {4 \p^2 \a ' \over \sqrt{V_M}} \sqrt{ g Q_1 \a '}\label{r1r5}
\eea
with $V_M$ the volume of $M$ in the metric $ds^2_{M}$ and $Q_1, \ Q_5$ the D1- and D5 charges.
The coordinates $\t, \o, \x$ vary over ${\bf R}$ while
$ \ 0 \leq \psi, \theta < \p, \ 0 \leq \varphi < 2 \p $.

\subsection{Killing spinors}
The D1-D5 background preserves 8 supersymmetries which get enhanced to 16 supersymmetries
in the near-horizon limit.
We will now derive
the explicit expression for the near-horizon Killing spinors needed in the following sections.

In type IIB supergravity, the supersymmetry variation parameter $\ve$
consists of two chiral spinors of the same chirality:
\be
\ve =
\left( \begin{array}{c}\ve_1 \\ \ve_2 \end{array} \right)
\ee
where $\G_{(10)} \ve_{1,2} = \ve_{1,2}$ with $ \G_{(10)} \equiv \G^0
\ldots \G^9$.
We now examine the conditions for $\ve$ to be a Killing spinor. Our
supergravity conventions follow \cite{Bena:2002kq} and the
dilatino and gravitino variations read
\footnote{Our 10D index conventions are as follows: $M,N = 0, \ldots 9,\ \m,\n = 0, \ldots, 5,\ m,n = 6, \ldots, 9$.
Hatted indices $\hat{M}, \hat{\m}, \hat{m}$ refer to a coordinate basis while unhatted ones $M,\m , m$ are
 orthonormal frame indices.}
\bea
\delta\lambda &=& -{1\over 4}e^\phi \sF{}_{(3)}\sigma^1\ve \nonu
\delta\Psi_{\hat {M}} &=& \nabla_{\hat{M}}\ve
+{e^\phi\over 8} \sF{}_{(3)}
\Gamma_{\hat{M}}\sigma^1 \ve\nonumber
\eea
The vanishing of the dilatino variation amounts to a chirality projection
\be \G_{(6)} \ve = \G_{(4)} \ve = - \ve  \label{cond1} \ee
where $ \G_{(6)} \equiv \G^0 \ldots \G^5$, $ \G_{(4)} \equiv \G^6 \ldots \G^9$.
while the gravitino variation with index on the internal manifold $M$
imposes that $\ve$ is covariantly constant in the internal directions:
\be  \nabla_{\hat{m}} \ve = 0. \label{cond2} \ee
The gravitino variation with index on $AdS_3 \times S^3$ then leads to
the equations
\be \d \Psi_{\hat{\m}} = \left[
\nabla_{\hat{\m}} + {1 \over 2 \sqrt{r_1 r_5}} \G^{012}\G_{\hat{\m}} \s^1 \right] \ve =0
\label{gravitino}\ee
Symmetry dictates that  solutions to this equation should be given by multiplying a constant spinor by
an $SL(2,R) \times SU(2)$ group element in a suitable representation \cite{Alonso-Alberca:2002gh}.
Expressing the spin connection on $AdS_3 \times S^3$ in terms of the following vielbein\footnote{Note that this is not the
left-invariant basis on the $SL(2,R) \times SU(2)$ group manifold but
rather a linear combination of left- and right invariant
forms.}
$$\begin{array}{ll}
e^0 = \sqrt{r_1 r_5} \cosh \x \cosh \o dt  & e^3 =  \sqrt{r_1 r_5} d \psi \\
e^1 = \sqrt{r_1 r_5} \cosh \x  d\o  & e^4 =  \sqrt{r_1 r_5} \sin \psi d \theta \\
e^2 = \sqrt{r_1 r_5} d\x  & e^5 =  \sqrt{r_1 r_5} \sin \psi \sin \theta d \f. \\
\label{vielbein}
\end{array}$$
one
finds the solutions
\be
\ve = e^{ {\x \over 2} \G^{02} \s^1} e^{ {\o \over 2} \G^{10} \s^1} e^{ {\t \over 2} \G^{21} \s^1}
 e^{ { {\p \over 2 } - \psi \over 2} \G^{45} \s^1} e^{ { {\p \over 2 } - \theta \over 2}
 \G^{35} \s^1} e^{ {\f \over 2} \G^{43} \s^1} \ve_0 \label{Killingspinor}
\ee
Here, $\ve$ is independent of the $AdS_3 \times S^3$ coordinates and satisfies the conditions
(\ref{cond1}), (\ref{cond2}):
\bea
\pa_{\hat{\m}} \ve_0 = \nabla_{\hat{m}} \ve_0 &=& 0 \nonu
 \G_{(6)} \ve_0 = \G_{(4)} \ve_0 &=& - \ve_0  \label{epsilonzero}
\eea

We can write a more explicit expression for $\ve_0$ by decomposing the
$SO(1,9)$ gamma matrices under the $SO(1,5) \times
SO(4)$ subgroup as follows:
\bea
\G^\m &=& \g^\m \otimes 1 \qquad \m = 0 \ldots 5\\
\G^m &=& \g_{(6)} \otimes \g^m \qquad m = 6 \ldots 9
\eea
where $\g^\m$ and $\g^m$ are $SO(1,5)$ and $SO(4)$ gamma matrices
respectively, and we have defined $\g_{(6)} = \g^0 \ldots \g^5$,
$\g_{(4)} = \g^6 \ldots \g^9$.
The ten-dimensional chirality operator is  $\G_{(10)} \equiv \G^0
\ldots \G^9 = \g_{(6)} \otimes \g_{(4)}$. A chiral spinor in ten
dimensions then decomposes as
$$ {\bf 16} \to {\bf (4,2)} + {\bf (4',2')}.$$ where the unprimed
(primed) representations have positive (negative) chirality. When $M$
is $K_3$, we take the convention that the representation ${\bf 2}$
forms a doublet under the SU(2) holonomy, while ${\bf 2'}$ consists of
two holonomy singlets. The chirality condition in (\ref{epsilonzero}) projects
out the ${\bf (4,2)}$ component. Choosing  basis elements $\h_+, \h_-$ for the
covariantly constant ${\bf 2'}$ spinors, we can take the following ansatz  for $\ve_0$:
\be
\ve_0 = \left( \begin{array}{c}\e_1^+  \\ \e_2^+
\end{array} \right) \otimes \h_+ + \left( \begin{array}{c}\e_1^- \\ \e_2^-
\end{array} \right)  \otimes \h_-.\label{ve0}\ee
with $\e^\pm$ constant and antichiral ($\g_{(6)}\e^\pm = - \e^\pm$) doublets on $AdS_3 \times S^3$  and $\h_\pm$
covariantly constant and antichiral ($\g_{(4)}\h_\pm = - \h_\pm$) spinors on $M$. Both for $M = T^4$ and $M= K3$,
 we have 16 independent Killing spinors.

\subsection{Poincar\'{e} Supersymmetries}
The following coordinate transformation takes us to Poincar\'{e} coordinates $(t, x , u)$ for $AdS_3$:
\bea
u &=& {1 \over \sqrt{r_1 r_5}} ( \cosh \x \cosh \o \cos \t + \cosh \x \sinh \o )\nonu
t &=& {1 \over u}  (\cosh \x \cosh \o \sin \t)  \nonu
x &=& {1 \over u} \sinh \x. \label{poinccoords}
\eea
The $AdS_3$ part of the metric and 3-form become
\bea ds^2_{AdS_3}&=& r_1 r_5 [ u^2 (-dt^2+dx^2) + {du^2 \over u^2} ]\nonu
F^{(3)}_{AdS_3}&=& {2r_5 ^2 \over g} u  dt \wedge dx \wedge du.\nonumber \eea
The 16 near-horizon Killing spinors split into 8 spinors that extend to the full asymptotically flat spacetime
(as they generate a Poincar\'{e} superalgebra we will henceforth refer to  them as `Poincar\'{e} supersymmetries')
and 8 spinors corresponding to enhanced near-horizon supersymmetries  (generating special conformal transformations).
In Poincar\'{e} coordinates, the Poincar\'{e} supersymmetries are time-independent  and are given by:
\be
\ve_P = \sqrt{u} R \ve_- \label{poincsusy}
\ee
where $R$ is the $SU(2) $ group element
$$ R = e^{ { {\p \over 2 } - \psi \over 2} \G^{45} \s^1} e^{ { {\p \over 2 } - \theta \over 2}
 \G^{35} \s^1} e^{ {\f \over 2} \G^{43} \s^1}$$
and $\ve_-$ is a spinor that satisfies, in addition to  (\ref{epsilonzero}), the extra projection condition
\be \G^{01} \s_1 \ve_- = - \ve_- . \label{poincproj} \ee
Here we have numbered the coordinates as $(x^0,x^1,x^2) = (t, x , u)$. See appendix \ref{appendix} for more
details on how the Poincar\'{e} supersymmetries extend to the full asymptotically flat geometry.

\subsection{Supersymmetric D-brane probes}

A supersymmetry of the background is preserved in the presence of a
bosonic Dp-brane configuration if it can be compensated for by a
$\k$-symmetry transformation \cite{Bergshoeff}. This can be expressed as a projection equation
\be (1 - \G ) \ve = 0 \label{kappaproj}\ee where $\G$ (satisfying ${\rm tr} \G = 0,\ \G^2 = 1$) is the operator entering in the $\k$-symmetry
transformation   rule on the Dp-brane and $\ve$ is a general Killing spinor (constructed above) pulled back to the
world-volume.
The operator $\G$ can be written in a simple form in a special worldvolume  Lorentz frame in which the
worldvolume field strength $F$ takes the form
$$ 2 \p \a' F = \tanh \F_0 e^{\underline{0}} \wedge e^{\underline{1}} + \sum_{r=1}^{(p-1)/2} \tan \F_r e^{\underline{2r}}
 \wedge e^{\underline{2r+1}}.$$
$\G$ is then given by \cite{Bergshoeff}
\begin{equation}
\label{iibgamma}
\G = e^{-a} \Gamma_{(0)} (\sigma_3)^{p-3\over2} i \sigma_2
\end{equation}
with
\bea
\Gamma_{(0)}&=&
\Gamma_{\underline{0} \ldots \underline{p}} \nonu
a &=& \sum_{r=0}^{(p-1)/2} \F_r \G^{\underline{2r} \underline{2r + 1}}\s_3.
\eea
In the above formulas, underlined indices are  orthonormal frame indices on the D-brane worldvolume.

\section{D-branes spanning $AdS_2$}\label{ads2branes}
In this section, we will consider D-branes that span an $AdS_2$ subspace within $AdS_3 \times S^3$. They
can be taken to be embedded at constant $\x =
\x_0$ in the coordinates (\ref{globalcoords}). We will see that the requirement of supersymmetry fixes $\x_0$
in terms of the charges carried by the brane.
\subsection{D1-brane along $AdS_2$}
\subsubsection{Near-horizon supersymmetries}
We consider a D1-brane probe embedded in $AdS_3$ at constant $\x =
\x_0$ and static in the remaining $S^3 \times M$ directions. The worldvolume coordinates can be taken to be
$\t, \o$.  We allow for an electric
field on the worldvolume which is conveniently parametrized as
$$ 2 \p \a ' F = \tanh \F_0 e^0 \wedge e^2.$$
Here and in what follows, with a slight abuse of notation, the
$e^a$ stand for the corresponding target space vielbein elements pulled-back to the world-volume. Explicitly, we
have
\bea
e^0 &=& \sqrt{r_1 r_5} \cosh \x_0 \cosh \o d \t \nonu
e^2 &=& \sqrt{r_1 r_5} \cosh \x_0 d\o \label{e0e1}
\eea
The supersymmetries preserved by the brane should satisfy $ (1 - \G ) \ve = 0$ with
$$\G = e^{- \F_0 \G^{02} \s^3} \G_{02} \s^1$$
$$ \ve = e^{ {\x_0 \over 2} \G^{20} \s^1} e^{ {\o \over 2} \G^{10}\s^1}
e^{ {\t \over 2} \G^{21} \s^1} R_0\ve_0 $$
where $R_0$ is a constant $SU(2)$ group element depending
on the position in the $S^3$ given by
$$R_0 \equiv R(\psi_0, \theta_0, \f_0) = e^{ { {\p \over 2 } - \psi_0 \over 2} \G^{45} \s^1} e^{ { {\p \over 2 } - \theta_0 \over 2}
 \G^{35} \s^1} e^{ {\f_0 \over 2} \G^{43} \s^1}.$$
Imposing $ (1 - \G ) \ve = 0$ for all
values of $\t, \o$ leads to two equations
\be
( 1 - s e^{- a}\G_{01} \s^1 ) e^{ s \frac{\xi_0}{2}\Gamma^{02}\sigma^1}
R_0 \ve_0 =0 \label{R0}
\ee
where $s = \pm 1$. Multiplying with $ e^{ s \frac{\xi_0}{2}\Gamma^{02}
\sigma^1}$ and taking linear combinations one finds that a  solution exists
if
$$ \tanh\x_0 = - {1 \over \cosh \F_0} \qquad \Leftrightarrow \qquad | \tanh \F_0 | = {1 \over \cosh \x_0 } .$$
Plugging in our ansatz for $\ve_0$  (\ref{ve0}), the projection condition on the surviving supersymmetries
can be written in terms of the 6-dimensional spinor doublet $\e^\pm$ as
\be
(1 - {\rm sgn} (\F_0)  R_0^{-1} \g_{02} \s^3 R_0 ) \e^\pm =0. \label{D1susys}
\ee
Hence the brane preserves half the supersymmetries of the background,
and the preserved supercharges depend on  the position of the brane on
$S^3$ through $R_0$ as well as on the sign of $\F_0$.

The latter is related to the sign of the fundamental string charge bound to the D1-brane.
Indeed, for nonzero $\F_0$, the D1-brane acts as a source for the B-field and carries an induced fundamental string
charge as well.
Demanding that it is properly quantized and equal to $q$ imposes a quantization condition on $\F_0$:
$$ \sinh \F_0 = {g r_1 \over  r_5 } q .$$
Note that, from (\ref{D1susys}), it follows that
branes carrying opposite fundamental string charge can preserve the
same supersymmetries provided they sit at antipodal locations on the
$S^3$. A similar property was observed for branes in other $AdS_p \times S^q$ backgrounds \cite{Simons:2004nm,Gaiotto:2006ns}.

The radial position $\x_0$ is determined by the fundamental string charge as
$$ \sinh \x_0 = { r_5 \over g r_1} {1 \over |q|}.$$
Of course, the above equations provide a solution to the equations of motion following from the DBI action
as one can easily verify.

The above $(q,1)$ string solution S-dualizes to a $(1,q)$ string the F1-NS5 background.
For $q=1$, the latter solution was found from the DBI equations of motion in \cite{Bachas}. Our analysis
implies that this solution should  be supersymmetric as well.

\subsubsection{Poincar\'{e} Supersymmetries}
We now check whether the above solution preserves any Poincar\'{e} supersymmetries.
A D1-brane at constant $\x = \x_0$ satisfies, in Poincar\'{e} coordinates (\ref{poinccoords}),
$$ u(x) = {\sinh \x_0 \over x} .$$
Taking $(t,x)$ to be the worldvolume coordinates, the $\k$-projector becomes
$$ \G = e^{- \F_0 \G^{\underline{0}\underline{1}} \s_3}\G_{\underline{0}\underline{1}}
\s_1 $$
with
\be\G_{\underline{0}\underline{1}} = - \tanh \x_0 \G_{01} + {1 \over \cosh \x_0}\G_{02}.\label{indgamma01}\ee
To check whether the D1-brane preserves some fraction  of the Poincar\'{e} supersymmetries, we need to verify whether
the equation $(1- \G) \ve_P$ (with $\ve_P$ given in (\ref{poincsusy})) has any solutions. Using
(\ref{poincproj})
one finds
the equation
$$
[ 1 + \cosh \F_0 \tanh \x_0 - {\cosh \F_0 \over \cosh \x_0 } \G_{02} \s_1 + \sinh \F_0 i \s_2 ] R \ve_- = 0
$$
As before, a solution exists when $\tanh \x_0 = -{1 \over \cosh \F_0}$ and requires
$$
(1 \pm R_0^{-1} \G_{02} \s_3 R_0)  \ve_- = 0.
$$
where the sign again depends on the sign of $\F_0$.
This projection condition is compatible with (\ref{poincproj}) and we conclude that the D1-brane preserves half
of the Poincar\'{e} supersymmetries.

\subsection{D3-branes along $AdS_2$ and wrapping a 2-cycle in $M$ }
\subsubsection{Near-horizon supersymmetries} \label{ads2t2section}
Here, we consider a D3-brane spanning and $AdS_2$ subspace in $AdS_3$ at $\x = \x_0$ and wrapping
a 2-cycle $\S$ in $M$. We denote the pull-back of the induced volume form on  $\S$ by ${\rm vol}_\S$
and define a corresponding $\G$-matrix combination:
\be \G_\S = {1 \over 2 \sqrt{g'}} \e^{\hat{a}\hat{b}} \G_{\hat{a}\hat{b}}\label{gammasigma}\ee
with $g'_{\hat{a}\hat{b}}$ the induced metric on $\S$.
We parametrize the worldvolume
flux as
$$ 2 \p \a ' F = \tanh \F_0 e^0 \wedge e^2 + \tan \F_1 {\rm vol}_\S $$
and take $\cos \F_1 \geq 0$.
The $\k$-projector is given by
$$ \G = e^{-\F_0 \G^{02} \s_3} e^{-\F_1 \G_\S \s_3} \G_{02}\G_\S i \s_2. $$
Imposing the supersymmetry condition $(1 - \G) \ve = 0$ leads to two equations
$$ ( 1 - e^{-\F_0 \G^{02} \s_3} e^{- s \F_1 \G_\S \s_3} \G_{02}\G_\S i \s_2 )
e^{s {\x_0 \over 2} \G^{02} \s_1} R_0 \ve_0 =0 $$
with $s = \pm 1$ and $R_0$ defined in (\ref{R0}).
After some manipulations  one finds that a solution exists if
\be \tanh \x_0 = -{\sin \F_1 \over  \cosh \F_0} \label{ads2t2sol} \ee
and requires the projection
$$\left( 1 - R_0^{-1} (  {\sinh \F_0 \over \cos \F_1} \G_\S \s_1 +
{ \cosh \F_0 \over  \cosh \x_0 \cos \F_1}  \G_{02}\G_\S i \s_2)R_0
\right) \ve_0 = 0$$
This projection equation can be
rewritten in a more standard form using the identity
$$ {\cosh \F_0 \over \cosh \x_0} = \sqrt{ \sinh ^2 \F_0 + \sin^2  \F_1} $$
which follows from (\ref{ads2t2sol}). The projection condition becomes
\be \left( 1 + R_0^{-1} e^{- {\a \over 2} \G^{02} \s_3} \G_{02}\G_\S i \s_2 e^{ {\a \over 2} \G^{02} \s_3} R_0 \right)
 \ve_0 = 0 \label{2cycleproj} \ee
with $\a$ defined by $$ \sinh \a = {\sinh \F_0 \over \cos \F_1  }.$$
In the above equation, both $\G_\S$ and $\ve_0$ are in general dependent on the position on $\S$ and it is not trivial that
the equation can be satisfied everywhere. This will possible  if $\S$ is a supersymmetric cycle. We proceed by
plugging
in our ansatz for $\ve_0$ (\ref{ve0}):
\be
\ve_0 = \left( \begin{array}{c}\e_1^+  \\ \e_2^+
\end{array} \right) \otimes \h_+ + \left( \begin{array}{c}\e_1^- \\ \e_2^-
\end{array} \right)  \otimes \h_-.\ee

In this ansatz, we are free to choose a convenient basis $\h_+, \h_-$ for the internal covariantly constant spinors.
It turns out that they
can be chosen to be eigenstates of $\G_\S$. Indeed, when $M = T^4$,
$\S$ is a $T^2$ within $T^4$
and
$\G_\S$ is position independent in suitable coordinates. The constant spinors $\h_+, \h_-$ can be chosen to
diagonalise $\G_\S$:
$ \G_\S \h_\pm =  \mp i \h_\pm.$

When $\S$ is a supersymmetric cycle in $M = K3$ and, we can also choose $\h_+, \h_-$ to diagonalise $\G_\S$.
Because $K3$ is hyperk\"{a}hler, it admits an $S^2$ family of complex structures,
and we assume $\S$ to be holomorphic with respect to one of these complex structures.
 Choosing holomorphic coordinates $z^i, \bar{z}^{\bar{i}}$  with respect to this particular complex structure
 we can choose a basis $\h_+, \h_-$  of covariantly constant spinors on $K3$ satisfying
\bea
\g_{\bar i} \h_+ = 0, & \g_{ij} \h_+ = \O_{ij} \h_- \nonu
\g_{i} \h_- = 0, & \g_{\bar{i}\bar{j}} \h_- = - \bar{\O}_{\bar i \bar j} \h_+ \label{covbasis}.
\eea
with $\O$ the $(2,0)$ form.
$\G_\S$ acts on $\h_\pm$ as
$\G_\S \h_\pm = \mp i \h_\pm .$

Summarizing, both on $M=T^4$ and $M= K3$ we can take $\h_+, \h_-$ to satisfy
$$ \G_\S \h_\pm =  \mp i \h_\pm.$$
Substituting into (\ref{2cycleproj}) gives projection conditions on the 6-dimensional spinor doublets
$ \e^+ = \left( \begin{array}{c}\e_1^+ \\ \e_2^+ \end{array} \right), \ \e^- = \left( \begin{array}{c}\e_1^- \\ \e_2^-
\end{array} \right)$ :
$$\left( 1 \pm R_0^{-1} e^{- {\a \over 2} \g^{02} \s_3} \g_{02}\s_2 e^{ {\a \over 2} \g^{02} \s_3} R_0 \right) \e^\pm
= 0$$
We see that indeed half of the supersymmetries is preserved.

\subsubsection{Poincar\'{e} supersymmetries}
It's straightforward to show that these branes preserve half of the Poincar\'{e} supersymmetries as well.
The projection condition on the Poincar\'{e} Killing spinors becomes
$$
\left( 1 - e^{ - \F_0 \G^{\underline{0}\underline{1}} \s_3} e^{- \F_1 \G_\S \s_3}\G_{\underline{0}\underline{1}}
 \G_\S i \s_2 \right) \ve_- = 0
$$
with $\G_{\underline{0}\underline{1}}$ defined in (\ref{indgamma01}).
Using the equation of motion (\ref{ads2t2sol}) and (\ref{poincproj})
this reduces to the projection equation
$$\left( 1 + R_0^{-1} e^{- {\a \over 2} \G^{02} \s_3} \G_{02}\G_\S i \s_2 e^{ {\a \over 2} \G^{02} \s_3} R_0 \right) \ve_- = 0 $$
with $\a$ defined by $$ \sinh \a =  {\sinh \F_0 \over \cos \F_1  }.$$
Note that this projection condition is compatible with (\ref{poincproj}) and can be solved as in the previous section
using the fact that $\S$ is a supersymmetric cycle. Hence we conclude that the D3-brane preserves half
of the Poincar\'{e} supersymmetries.

\subsection{D5-branes spanning $AdS_2 \times M$}

\subsubsection{Near-horizon supersymmetries}

In this subsection we consider a D5-brane spanning and $AdS_2$ subspace in $AdS_3$ at $\x = \x_0$ and wrapping the
whole of $M$. Choosing suitable complex coordinates on $M$ we can take the
 the worldvolume
flux as
$$ 2 \p \a ' F = \tanh \F_0 e^0 \wedge e^2 + i \tan \F_1 e^1 \wedge e^{\bar{1}}  + i \tan \F_2 e^2 \wedge e^{\bar{2}}  $$
with $\cos \F_{1,2} \geq 0$.
The $\k$-projector is given by
$$ \G = e^{-\F_0 \G^{02} \s_3} e^{- i \F_1 \G^{1\bar{1}} \s_3} e^{-i \F_2 \G^{2\bar{2}} \s_3} \G_{02}\G_{(4)}  \s_1 .$$
Requiring $(1 - \G) \ve = 0$  and using the chirality property (\ref{epsilonzero}) leads to two equations
$$ ( 1 + s e^{-\F_0 \G^{02} \s_3} e^{- i s (\F_1 + \F_2 ) \G^{1\bar{1}} \s_3} \G_{02}  \s_1 )
e^{s {\x_0 \over 2} \G^{02} \s_1} R_0 \ve_0 =0 $$
with $s = \pm 1$.
Note that only the sum $\F \equiv \F_1  + \F_2$ of the worldvolume flux parameters on $M$ enters the equations,
while their difference is
left undetermined.
After some algebra, one finds that a solution exists if
\be \tanh \x_0 =  {\cos  \F  \over  \cosh \F_0}.\label{ads2t4sol}\ee
and requires the projection
$$\left( 1 + R_0^{-1} e^{- {\a \over 2} \G^{02} \s_3} \G_{02}\G_{1\bar{1}} \s_2 e^{ {\a \over 2} \G^{02} \s_3} R_0
\right) \ve_0 = 0 $$
where we defined  $\a$  by $$ \sinh \a = {\sinh \F_0 \over \sin \F }.$$ We proceed by plugging
in the ansatz for $\ve_0$ (\ref{ve0}) and choosing internal spinors $\h_+ , \ \h_-$ satisfying
$$ \G_{1\bar{1}} \h_\pm  = \mp \h_\pm.$$ This can trivially done for $M=T^4$, while for $M = K3$ one
can choose $\h_\pm$ to obey (\ref{covbasis}). The resulting 6D projection conditions
$$\left( 1 \pm R_0^{-1} e^{- {\a \over 2} \G^{02} \s_3} \G_{02}\G_{1\bar{1}} \s_2 e^{ {\a \over 2} \G^{02} \s_3} R_0
\right) \e^\pm = 0 $$
show that the brane is half-BPS.

\subsubsection{Poincar\'{e} supersymmetries}

As in the previous cases, these branes preserve half of the Poincar\'{e} supersymmetries as well.
The $\k$-projection condition on the Poincar\'{e} Killing spinors (\ref{poincsusy}) leads to a single equation
$$
\left( 1 + e^{ - \F_0 \G_{\underline{0}\underline{1}} \s_3} e^{- i \F \G^{1\bar{1}}\s_3}
\G_{\underline{0}\underline{1}}\s_1 \right) \ve_- = 0
$$
with
 $\G_{\underline{0}\underline{1}}$ defined in (\ref{indgamma01}).
Using the equation of motion (\ref{ads2t4sol})
 and (\ref{poincproj})
this reduces to  the projection equation
$$\left( 1 + R_0^{-1} e^{- {\a \over 2} \G^{02} \s_3} \G_{02}\G_{1\bar{1}} \s_2 e^{ {\a \over 2} \G^{02} \s_3} R_0 \right) \ve_- = 0 $$
with $\a$ again defined by $$ \sinh \a = {\sinh \F_0 \over \sin \F }$$
This projection condition is compatible with (\ref{poincproj}) and we conclude that the D5-brane preserves half
of the Poincar\'{e} supersymmetries.

\section{D-branes spanning $AdS_2 \times S^2$}\label{ads2s2branes}
In this section we consider D-branes spanning an $AdS_2 \times S^2$ subspace within $AdS_3 \times S^3$.
They can be taken to be embedded at constant $\x =\x_0 $ and $\psi = \psi_0$ in the coordinate system (\ref{globalcoords}).

\subsection{D3-branes along $AdS_2 \times S^2$}\label{ads2s2branes1}
\subsubsection{Near-horizon supersymmetries}
We
consider a D3-brane probe  in this background sitting at constant $\x =
\x_0$ and $\psi = \psi_0$ and static on $M$.  The worldvolume coordinates can be taken to be
$(\t,\omega,\theta,\phi)$ and we allow for an electromagnetic field
 on the worldvolume  parametrized as
$$ 2 \p \a ' F = \tanh \F_0 e^0 \wedge e^2 + \tan \F_1 e^4 \wedge e^5
$$
with $\cos \F_1 \geq 0$, $e^0, \ e^1$ as in (\ref{e0e1}) and
\bea
e^4 &=& \sqrt{r_1 r_5} \sin \psi_0 d \theta \nonu
e^5 &=& \sqrt{r_1 r_5} \sin \psi_0 \sin \theta d \varphi \label{e4e5}
\eea
Then the supersymmetries preserved by the brane are solutions of $
(1 - \G ) \ve = 0$ where the Killing spinor $\ve$ is given in
(\ref{Killingspinor}), (\ref{epsilonzero}) and
$$\G =   e^{-  \F_0 \G^{02} \s^3 } e^{- \F_1 \G^{45} \s^3}\G_{0245} i\s^2.$$
Imposing $ (1 - \G ) \ve = 0$ for all values of
$\t, \o, \theta , \f$ leads to four equations
\be
[ 1 - e^{ - s_2 \F_0  \G^{02} \s^3} e^{ - s_1 \F_1  \G^{45} \s^3}
i \G_{0245} \s^2 ]
e^{ s_1 \frac{\xi_0}{2}\G^{02} \sigma^1} e^{ s_2 {{\p \over 2} - \psi_0 \over 2}
\G^{45} \sigma^1}
\ve_0 =0
\ee
with $s_{1,2} = \pm 1$. Multiplying with $e^{s_1
\frac{\xi_0}{2}\G^{02} \sigma^1} e^{ s_2 {{\p \over 2} - \psi_0 \over 2}
\G^{45} \sigma^1} $ and taking linear combinations one finds
that solution exist if
\bea
\tanh \x_0 &=& -{ \sin \F_1 \over  \cosh \F_0 } \nonu
\cot \psi_0 &=& - { \sinh \F_0 \over \cos \F_1} \label{ads2s2sol}
\eea
which implies that the worldvolume fluxes take the values
\bea
\tanh \F_0 &=& -{\cos \psi_0 \over  \cosh \x_0}\nonu
\tan \F_1 &=& -{\sinh \x_0 \over  \sin \psi_0}\nonumber
\eea
and the preserved supersymmetries have to  satisfy
\be
(1 - \G_{0245} i \s^2 ) \ve_0 = 0.
\ee
Plugging in our ansatz for $\ve_0$  (\ref{ve0}), this
can be written in terms of the 6-dimensional spinor doublets $\e^\pm$ as
\be
(1 - \g_{0245} i \s^2 ) \e^\pm = 0.\label{ads2s2proj}
\ee
Hence such D3-branes are half-BPS.

In the presence of electric and magnetic worldvolume flux the D3 brane sources the electric NS and R
two forms $B^{(2)}$ and $C^{(2)}$ and
carries induced $F-$ and $D$-string charges.  This imposes two charge
quantization conditions which can be computed from requiring that the
solution provides sources for $B^{(2)}$ and $C^{(2)}$ with quantized
coefficients. The quantization conditions read:
\bea
\psi_0 = {g \p \a ' \over r_5^2} q &=& {q \over Q_5}\p \label{Fcharge}\nonu
\sinh \x_0 \sin \psi_0 &=& { \p \a ' \over r_1 r_5} p\label{Dcharge}
\eea
where $q,p$ are the induced $F-$ and $D$-string charges respectively and we have used (\ref{r1r5}). Since
$0\leq\psi_0\leq\p$ we see that the radius of the $S^2$ can take
on $Q_5$ different values. All these solutions preserve the same set
of supersymmetries as follows from (\ref{ads2s2proj}). Since the $S^2$ in $S^3$ is contractible, these branes do
not carry a net $D3$ charge and should be most likely interpreted as bound states of $(p,q)$ strings `puffed-up' through a version
of the Myers effect \cite{Myers:1999ps}.

The S-dual version of this solution in the F1-NS5 background was constructed and shown to be supersymmetric  in
\cite{Bachas}. Note that our solution confirms the charge quantization conditions found there which are more subtle
in the S-dual background due to the presence of background NS flux.

\subsubsection{Poincar\'{e} Supersymmetries} \label{ads2s2poinc}
We can again check whether the solution preserves any Poincar\'{e} supersymmetries.
For a D3-brane spanning an $AdS_2 \times S^2$ subspace $\x = \x_0, \psi = \psi_0$ in global coordinates,
the $\k$-projector in Poincar\'{e} coordinates takes the form
$$ \G =  e^{- \F_0 \G^{\underline{0}\underline{1}} \s_3} e^{- \F_1 \G^{45} \s_3 }
 \G_{\underline{0}\underline{1}45} i  \s_2.
$$
with $\G_{\underline{0}\underline{1}}$ defined in (\ref{indgamma01}).
Requiring $(1- \G) \ve_P=0$ with $\ve_P$ given in (\ref{poincsusy}), (\ref{poincproj}) for all values of $\theta, \varphi$ on $S^2$ leads to two equations
\be
\left( 1 - e^{ - s \F_0  \G^{\underline{0}\underline{1}} \s^3} e^{ -  \F_1  \G^{45} \s^3}
 \G_{\underline{0}\underline{1}45} i \s^2\right)
 e^{ s {{\p \over 2} - \psi_0 \over 2}
\G^{45} \sigma^1}
\ve_- =0
\ee
with $s = \pm 1$. Multiplying by $ e^{ s {{\p \over 2} - \psi_0 \over 2}} e^{   \F_1  \G^{45} \s^3}$ and
using (\ref{poincproj})  and the equations of motion (\ref{ads2s2sol})
one gets a single condition
$$ (1 - \G_{0245} i \s_2 ) \ve_- =0.$$
This is consistent with (\ref{poincproj}) and again we see that half of the Poincar\'{e} supersymmetries are preserved.

\subsection{D5-branes spanning $AdS_2 \times S^2$ and wrapping a 2-cycle in $M$}
Next, we consider D5-branes that span and $AdS_2 \times S^2$ subspace and wrap
a 2-cycle $\S$ in $M$. Although one can construct solutions to the DBI equations of this form,
none of them is actually supersymmetric as we will presently show.

Parametrizing the worldvolume
flux as
$$ 2 \p \a ' F = \tanh \F_0 e^0 \wedge e^2 + \tan \F_2 e^4 \wedge e^5 + \tan \F_2 {\rm vol}_\S   $$
with $e^0,e^2, e^4, e^5$ as in (\ref{e0e1}), (\ref{e4e5})  the $\k$ projector is given by
$$ \G = e^{-\F_0 \G^{02} \s_3} e^{-\F_1 \G^{45} \s_3}  e^{-\F_2 \G_\S \s_3} \G_{024567} \s_1 $$
with $\G_\S$ as in (\ref{gammasigma}).
Requiring $(1 - \G) \ve = 0$ everywhere  leads to four equations
$$ \left( 1 - s_1 s_2 e^{- s_2 \F_0 \G^{02} \s_3 - s_1 \F_1 \G^{45} \s_3} e^{- s_1 s_2 \F_2  \G_{\S} \s_3} \G_{0245}\G_\S
   \s_1 \right)
e^{s_1 {\x_0 \over 2} \G^{02} \s_1 + s_2 {\p/2 - \psi_0 \over 2} \G^{45} \s_1} \ve_0 =0 $$
with $s_{1,2} = \pm 1$. Multiplying by
$e^{-s_1 {\x_0 \over 2} \G^{02} \s_1- s_2 {\p/2 - \psi_0 \over 2} \G^{45} \s_1}e^{ s_2 \F_0 \G^{02} \s_3+
 s_1 \F_1 \G^{45} \s_3+  s_1 s_2 \F_2  \G_\S \s_3}$ the four equations can be written out schematically as
$$ A(s_1,s_2) \ve_0 = \left( B(s_1,s_2) \s_1 + C(s_1,s_2) i\s_2 + D(s_1,s_2) \s_3 \right) \ve_0$$
where the coefficients $A,B,C,D$ don't depend on the $\s$-matrices. Anticommuting the $s_1 = 1, s_2 = 1$ and
$s_1 = 1, s_2 = -1$ equations leads to
$$( 1 + \sin^2 \psi_0  \cosh 2 \F_0  \cos 2 \F_2 +     \cos^2 \psi_0 \cos 2 \F_1) \ve_0 =
   - \sin^2 \psi_0 \sinh 2 \F_0  \sin 2 \F_2 \G^{02}\G_\S  \ve_0  $$
Since $(\G^{02}\G_\S)^2 = -1$, solutions are possible if both sides vanish separately. In particular,
one needs either $\sin \psi_0,\ \F_0$ or $\sin 2 \F_2$ to vanish. We found none of these cases to be consistent
with the remaining equations.

\subsection{D7-branes spanning $AdS_2 \times S^2 \times M$}
\subsubsection{Near-horizon supersymmetries}
Here we consider a D7-brane spanning and $AdS_2 \times S^2 $ subspace in $AdS_3 \times S^3 $ at $\x = \x_0,\ \psi = \psi_0$
and wrapping the whole of  $M$. Choosing complex coordinates on $M$,
the worldvolume flux can be brought in the form
$$ 2 \p \a ' F = \tanh \F_0 e^0 \wedge e^2 + \tan \F_2 e^4 \wedge e^5 + i \tan \F_2 e^1 \wedge e^{\bar{1}}  + i
\tan \F_3  e^2 \wedge e^{\bar{2}}  $$
and the $\k$ projector is given by
$$ \G = e^{-\F_0 \G^{02} \s_3} e^{-\F_1 \G^{45} \s_3}  e^{-i \F_2 \G^{1\bar{1}} \s_3} e^{-\F_3 \G^{2\bar{2}} \s_3}
\G_{0245}\G_{(4)} i \s_2 .$$
Requiring $(1 - \G) \ve = 0$  and using the chirality property (\ref{epsilonzero}) leads to four equations
\be ( 1 + e^{- s_2 \F_0 \G^{02} \s_3}  e^{- s_1 \F_1 \G^{45} \s_3} e^{- i s_1 s_2 \F \G^{1\bar{1}} \s_3} \G_{0245}  i \s_2 )
e^{s_1 {\x_0 \over 2} \G^{02} \s_1} e^{s_2 {\p/2 - \psi_0 \over 2} \G^{45} \s_1} \ve_0 =0 \label{ads2s2Meqs} \ee
with $s_1, s_2 = \pm 1$. Note that only the sum $\F \equiv \F_2 + \F_3$ enters the equations while the difference
is unconstrained. Even though solutions of the D-brane Born-Infeld equations exist for general $\F$, manipulations
similar to the ones in the previous section show that
that the  above supersymmetry conditions are consistent only for
$$ \F = 0.$$
Note that this implies that the worldvolume field strength on $M$ is anti-selfdual.
In this case, the equations (\ref{ads2s2Meqs}) reduce (up to a sign difference) to the ones
solved in section \ref{ads2s2branes1}. When $M=T^4$, this was to be expected from T-duality, which leaves the
background invariant and relates
the probe solutions. The solution is given by
\bea
\tanh \x_0 &=& { \sin \F_1 \over \cosh \F_0 } \nonu
\cot \psi_0 &=& - { \sinh \F_0 \over \cos \F_1 } \nonu
(1 - \g_{0245} i \s^2 ) \e^\pm &=& 0.
\eea
Comparing with (\ref{ads2s2proj}), we note that the $S^2$-wrapping D3-branes and D7-branes are mutually BPS.

As before, the values of $\xi_0, \psi_0$ are quantized in terms of the induced charges carried by the brane.
For nonzero $\F_0, \F_1$, the D7-brane provides a source for the NSNS 2-form $B^{(2)}$ and the D5-brane RR
potential $C^{(6)}$ and carries induced F1- and D5- charge. This leads to quantization conditions
\bea
\psi_0   &=& {q \over Q_1}\p \nonu
\sinh \x_0 \sin \psi_0 &=& { \p \a ' \over r_1 r_5} p_5
\eea
where $q$ and $p_5$ denote the induced F1- and D5- charge respectively. We see that, in this case, the
$S^2$ radius can take on $Q_1$ different values, and the corresponding solutions preserve the same set of supersymmetries.
\subsubsection{Poincar\'{e} supersymmetries}
A calculation almost identical to paragraph \ref{ads2s2poinc} shows that these branes preserve half of the
Poincar\'{e} supersymmetries as well.

\section{Discussion and outlook}\label{discussion}

In this paper, we have constructed a variety of supersymmetric probe brane solutions in the near-horizon
D1-D5 background. They are all static with respect to global time and preserve half of the near-horizon supersymmetries.
Since the global time generator corresponds to $L_0 + \bar{L}_0$ in the dual CFT, we expect these branes to
correspond to supersymmetric conformal operators in the dual CFT.
 In appendix \ref{otherbranes} we consider branes spanning some other submanifolds,
none of which is found to be supersymmetric.

As was mentioned in the Introduction, one of the motivations for studying
the branes constructed in this paper is the fact that they share some properties with  brane probes
in other D-brane backgrounds
that have been related to microstates
\cite{Gaiotto:2006ns,Gaiotto:2004ij,Kim:2005yb,Gaiotto:2006wm}.
An important open question is therefore whether some of the
D-branes considered here can be related to  the microstates of the D1-D5 system.
One way to clarify their role
would be to study their interpretation
from the point of view of the dual CFT description of the D1-D5 system (see \cite{Justin} for a review).
Branes spanning an $AdS_2$ subspace run off to the boundary of $AdS_3$ where they form a line defect in the dual CFT \cite{DeWolfe:2001pq,Bachas:2001vj}.
Similar $AdS_2$ branes in the $AdS_5 \times S^5$ background were given a dual CFT interpretation in \cite{Gomis:2006sb}.
We leave this interesting topic for further study.
A related issue concerns the relation, if any, of the probe brane solutions considered here and
 the microstate
geometries for the D1-D5 system \cite{Mathur}.

It would also be of interest to extend these solutions to the full  asymptotically flat geometry.
It may be mentioned that in the context of two dimensional black holes, branes with similar properties (in particular
in the asymptotically flat geometry) have been discussed \cite{KPY, Nakayama}.


\acknowledgments{We would like to thank Dileep Jatkar for initial collaboration and Yu Nakayama, Sumit Das,
Samir Mathur and  Gautam Mandal  for useful discussions. We would also like thank our Referee for useful comments.
KPY would like to gratefully acknowledge the hospitality of KIAS, Seoul  and H.R.I, Allahabad, India during the course
of this work. We would also like to thank the organisers of Strings 2006, Beijing for a wonderful conference and for
the opportunity to discuss our work. This work was supported by the Science Research Center Program of
the Korea Science and Engineering Foundation through
the Center for Quantum Spacetime(CQUeST) of
Sogang University with grant number R11 - 2005 - 021. }

\begin{appendix}
\section{Supersymmetries of the asymptotically flat background}\label{appendix}
In this appendix we sketch how the Poincar\'{e}e supersymmetries  (\ref{poincsusy}, \ref{poincproj}) arise from
the Killing spinors of the full asymptotically flat geometry.
We will use the solution for
the D1/D5 system given in e.g. \cite{Justin}
The dilatino equation
$$
(\G_M \del^M\F+ \G^{MNP} F_{MNP} ^{(3)}\s^1) \ve= 0
$$
becomes
$$
(-\frac{r_1 ^2}{f_1 r^3}+\frac{r_5 ^2}{f_5 r^3})\G^2 \ve -(\frac{r_1
^2}{f_1 r^3}\G^{012}+\frac{r_5 ^2}{f_5 r^3}\G^{345})\s^1\ve=0
$$
In order to have a solution, we need to impose
two projection conditions
\be
\G^{01}\s^1\ve=- \ve;  \qquad \G^{2345}\s^1\ve=\ve
\ee
The second equation can be traded for $\G_{(6)} \ve = - \ve$ and taking the near-horizon limit we recover
the two projection conditions (\ref{poincproj}, \ref{cond1}) imposed on  the Poincar\'{e}
supersymmetries.

In the near horizon limit however,  the first two terms in the above
dilatino equation cancel leaving behind a single projection condition equivalent to
\be
\G_{(6)} \ve = - \ve
\ee
which becomes the projection condition (\ref{cond1}) on the near-horizon
supersymmetries.

\section{D-branes spanning other submanifolds}\label{otherbranes}

In the main text, we have considered D-branes which span an $AdS_2$ subspace in $AdS_3$. In this appendix we explore some
other possibilities. We restrict attention to D-branes which are static with respect to global time. We find that none of
the branes considered here preserve any supersymmetry.

\subsection{D3-brane wrapping $S^3$}
Consider a static D3-brane at $\x = \x_0,\ \o = \o_0$ in $AdS_3$ and wrapping the $S^3$. The worldvolume gauge field can be brought in the form
 $$ 2 \p \a ' F = \tanh \F_0 e^0 \wedge e^3 + \tan \F_2 e^4 \wedge e^5 .$$
One easily checks that the DBI equations of motion impose $\F_0 = \x_0 = \o_0 =0$.  The conditions for such a  solution to preserve supersymmetry
are
$$ ( 1- s_1 e^{- s_1 s_2 \F_1 \G^{45} \s_3 } \G_{0345} i \s_2 ) \ve_0 =0$$
for $s_1, \ s_2 = \pm 1$. One easily checks that the resulting four equations cannot be solved simultaneously;
hence there are no supersymmetric  solutions in this case.

\subsection{Branes spanning $AdS_2$ wrapping a $T^2$ within $S^3$}
In this section, we will discuss branes wrapping a $T^2\in S^3$.  One motivation to consider such branes
(apart from an exercise of imagination) is from the viewpoint of the R-symmetry of the SCFT dual to the $AdS_3$
string theory.
Extended objects in $S^3$ will transform non-trivially under the $SO(4)$ isometry  of $S^3$ which becomes the
R-symmetry of the Higgs branch. The Coulomb branch has a different R-symmetry.  It is interesting to consider
the supersymmetry properties of such toroidal branes in $S^3$ then because such branes do not transform in some
obvious manner under $SO(4)$ (in contrast to branes which wrap an $S^2$ which are conjugacy classes of $SU(2)$ and
hence invariant under an vectorial $SU(2)$ of $SO(4)$).

In order to study these branes, we will use Euler-angle co-ordinates on $S^3$ - the metric takes the form
\be
d\theta^2  +\sin^2\theta d\f_1^2 +\cos^2\theta d\f_2^2
\ee
The spin-connection one form can be taken to be
$$
\o_{46}=\cos\theta d\f_1  \hspace{1cm} \o_{56}=-\sin\q d\f_2
$$
with all other components vanishing. Solving the gravitino equations the sphere part of the killing spinor
$$
e^{\frac{\q}{2}\G_{45}\s^1} e^{-\frac{\f_1}{2}\G_{46}} e^{-\frac{\f_2}{2}\G_{46}\s^1} \ve_0
$$ with $\ve_0$ a constant spinor. Note that the $4=\f_1, 5=\f_2,\q=6$.

The $\kappa$ symmetry matrix is $$\G=e^{-\frac{s^3}{2}(\F_0\G_{02}+\F_1\G_{45})}  \G_{0245} i\s^2$$
As before, we demand that $\G\ve=\ve$ for all $\t,\o,\f_1,\f_2$ and its a simple matter to check that this gives equations which cannot be satisfied (so long as $\x_0$ is finite i.e., the brane is at a finite radius away from the boundary). More specifically, the incompatible equations are those which come from imposing
$\G\G_{45}\ve=\G_{45}\ve$ and $\G\G_{45}\s^1 \ve=\G_{45}\s^1\ve$ (the latter two are the conditions which come from requiring that the kappa symmetry condition hold for all $\f_{1,2}$).

\end{appendix}


\begin{thebibliography}{10}

\bibitem{Simons:2004nm}
  A.~Simons, A.~Strominger, D.~M.~Thompson and X.~Yin,
  ``Supersymmetric branes in AdS(2) x S**2 x CY(3),''
  Phys.\ Rev.\ D {\bf 71}, 066008 (2005)
  [arXiv:hep-th/0406121].




\bibitem{Gaiotto:2006ns}
  D.~Gaiotto, A.~Strominger and X.~Yin,
  ``From AdS(3)/CFT(2) to black holes / topological strings,''
  arXiv:hep-th/0602046.



\bibitem{Li:2006uq}
  W.~Li and A.~Strominger,
  ``Supersymmetric probes in a rotating 5D attractor,''
  arXiv:hep-th/0605139.


\bibitem{Gaiotto:2004pc}
  D.~Gaiotto, A.~Simons, A.~Strominger and X.~Yin,
  ``D0-branes in black hole attractors,''
  arXiv:hep-th/0412179.

\bibitem{Rodriguez-Gomez:2005na}
  D.~Rodriguez-Gomez,
  ``Branes wrapping black holes as a purely gravitational dielectric effect,''
  JHEP {\bf 0601}, 079 (2006)
  [arXiv:hep-th/0509228].

\bibitem{Gaiotto:2004ij}
  D.~Gaiotto, A.~Strominger and X.~Yin,
  ``Superconformal black hole quantum mechanics,''
  JHEP {\bf 0511}, 017 (2005)
  [arXiv:hep-th/0412322].


\bibitem{Kim:2005yb}
  S.~Kim and J.~Raeymaekers,
  ``Superconformal Quantum Mechanics Of Small Black Holes,''
  JHEP {\bf 0508}, 082 (2005)
  [arXiv:hep-th/0505176].

\bibitem{Gaiotto:2006wm}
  D.~Gaiotto, A.~Strominger and X.~Yin,
  ``The M5-brane elliptic genus: Modularity and BPS states,''
  arXiv:hep-th/0607010.

\bibitem{Aharony:1999ti}
  O.~Aharony, S.~S.~Gubser, J.~M.~Maldacena, H.~Ooguri and Y.~Oz,
  ``Large N field theories, string theory and gravity,''
  Phys.\ Rept.\  {\bf 323}, 183 (2000)
  [arXiv:hep-th/9905111].

\bibitem{Justin}
  J.~R.~David, G.~Mandal and S.~R.~Wadia,
  ``Microscopic formulation of black holes in string theory,''
  Phys.\ Rept.\  {\bf 369}, 549 (2002)
  [arXiv:hep-th/0203048].

\bibitem{GG}
 J.~McGreevy, L.~Susskind and N.~Toumbas,
  ``Invasion Of The Giant Gravitons From Anti-De Sitter Space,''
  JHEP {\bf 0006} (2000) 008
  [arXiv:hep-th/0003075].


  B.~Janssen, Y.~Lozano and D.~Rodriguez-Gomez,
  ``Giant gravitons in AdS(3) x S**3 x T**4 as fuzzy cylinders,''
  Nucl.\ Phys.\ B {\bf 711} (2005) 392
  [arXiv:hep-th/0406148].

\bibitem{Das}
  S.~R.~Das, S.~Giusto, S.~D.~Mathur, Y.~Srivastava, X.~Wu and C.~Zhou,
  ``Branes wrapping black holes,''
  Nucl.\ Phys.\ B {\bf 733}, 297 (2006)
  [arXiv:hep-th/0507080].

\bibitem{Bachas:2000ik}
  C.~Bachas, M.~R.~Douglas and C.~Schweigert,
  ``Flux stabilization of D-branes,''
  JHEP {\bf 0005}, 048 (2000)
  [arXiv:hep-th/0003037].

\bibitem{Pawelczyk:2000hy}
  J.~Pawelczyk and S.~J.~Rey,
  ``Ramond-Ramond flux stabilization of D-branes,''
  Phys.\ Lett.\ B {\bf 493}, 395 (2000)
  [arXiv:hep-th/0007154].

\bibitem{Couchoud:2003jw}
  N.~Couchoud,
  ``Anti-de Sitter branes with Neveu-Schwarz and Ramond-Ramond backgrounds,''
  JHEP {\bf 0303}, 007 (2003)
  [arXiv:hep-th/0301195].

\bibitem{Bachas}
  C.~Bachas and M.~Petropoulos,
  ``Anti-de-Sitter D-branes,''
  JHEP {\bf 0102}, 025 (2001)
  [arXiv:hep-th/0012234].

  \bibitem{Bachas:2002nz}
P.~M.~Petropoulos and S.~Ribault,
  ``Some remarks on anti-de Sitter D-branes,''
  JHEP {\bf 0107}, 036 (2001)
  [arXiv:hep-th/0105252].\\
  A.~Giveon, D.~Kutasov and A.~Schwimmer,
  ``Comments on D-branes in AdS(3),''
  Nucl.\ Phys.\ B {\bf 615}, 133 (2001)
  [arXiv:hep-th/0106005].\\
  P.~Lee, H.~Ooguri, J.~W.~Park and J.~Tannenhauser,
  ``Open strings on AdS(2) branes,''
  Nucl.\ Phys.\ B {\bf 610}, 3 (2001)
  [arXiv:hep-th/0106129].\\
  Y.~Hikida and Y.~Sugawara,
  ``Boundary states of D-branes in AdS(3) based on discrete series,''
  Prog.\ Theor.\ Phys.\  {\bf 107}, 1245 (2002)
  [arXiv:hep-th/0107189].\\
  A.~Rajaraman and M.~Rozali,
  ``Boundary states for D-branes in AdS(3),''
  Phys.\ Rev.\ D {\bf 66}, 026006 (2002)
  [arXiv:hep-th/0108001].\\
  P.~Lee, H.~Ooguri and J.~w.~Park,
  ``Boundary states for AdS(2) branes in AdS(3),''
  Nucl.\ Phys.\ B {\bf 632}, 283 (2002)
  [arXiv:hep-th/0112188].\\
    C.~Bachas,
  ``Asymptotic symmetries of AdS(2) branes,''
  arXiv:hep-th/0205115.


\bibitem{Camino:2001at}
  J.~M.~Camino, A.~Paredes and A.~V.~Ramallo,
  ``Stable wrapped branes,''
  JHEP {\bf 0105}, 011 (2001)
  [arXiv:hep-th/0104082].

\bibitem{Skenderis:2002vf}
  K.~Skenderis and M.~Taylor,
  ``Branes in AdS and pp-wave spacetimes,''
  JHEP {\bf 0206}, 025 (2002)
  [arXiv:hep-th/0204054].

\bibitem{Drukker:2005kx}
  N.~Drukker and B.~Fiol,
  ``All-genus calculation of Wilson loops using D-branes,''
  JHEP {\bf 0502}, 010 (2005)
  [arXiv:hep-th/0501109].

\bibitem{Gomis:2006sb}
  J.~Gomis and F.~Passerini,
  ``Holographic Wilson loops,''
  JHEP {\bf 0608}, 074 (2006)
  [arXiv:hep-th/0604007].



\bibitem{Bachas:2001vj}
  C.~Bachas, J.~de Boer, R.~Dijkgraaf and H.~Ooguri,
  ``Permeable conformal walls and holography,''
  JHEP {\bf 0206}, 027 (2002)
  [arXiv:hep-th/0111210].







\bibitem{Maldacena:1997re}
  J.~M.~Maldacena,
  ``The large N limit of superconformal field theories and supergravity,''
  Adv.\ Theor.\ Math.\ Phys.\  {\bf 2}, 231 (1998)
  [Int.\ J.\ Theor.\ Phys.\  {\bf 38}, 1113 (1999)]
  [arXiv:hep-th/9711200].



\bibitem{Maldacena:1998bw}
  J.~M.~Maldacena and A.~Strominger,
  ``AdS(3) black holes and a stringy exclusion principle,''
  JHEP {\bf 9812}, 005 (1998)
  [arXiv:hep-th/9804085].

  \bibitem{Bena:2002kq}
  S.~F.~Hassan,
  ``T-Duality, Space-Time Spinors And R-R Fields In Curved Backgrounds,''
  Nucl.\ Phys.\ B {\bf 568} (2000) 145
  [arXiv:hep-th/9907152].

  I.~Bena and R.~Roiban,
  ``Supergravity pp-wave solutions with 28 and 24 supercharges,''
  Phys.\ Rev.\ D {\bf 67}, 125014 (2003)
  [arXiv:hep-th/0206195].

\bibitem{Alonso-Alberca:2002gh}
  N.~Alonso-Alberca, E.~Lozano-Tellechea and T.~Ortin,
   ``Geometric construction of Killing spinors and supersymmetry algebras in
  homogeneous spacetimes,''
  Class.\ Quant.\ Grav.\  {\bf 19}, 6009 (2002)
  [arXiv:hep-th/0208158].



\bibitem{Bergshoeff}
  E.~Bergshoeff and P.~K.~Townsend,
  ``Super D-branes,''
  Nucl.\ Phys.\ B {\bf 490}, 145 (1997)
  [arXiv:hep-th/9611173].

  E.~Bergshoeff, R.~Kallosh, T.~Ortin and G.~Papadopoulos,
  ``kappa-symmetry, supersymmetry and intersecting branes,''
  Nucl.\ Phys.\ B {\bf 502}, 149 (1997)
  [arXiv:hep-th/9705040].

  M.~Marino, R.~Minasian, G.~W.~Moore and A.~Strominger,
  ``Nonlinear instantons from supersymmetric p-branes,''
  JHEP {\bf 0001}, 005 (2000)
  [arXiv:hep-th/9911206].



\bibitem{Myers:1999ps}
  R.~C.~Myers,
  ``Dielectric-branes,''
  JHEP {\bf 9912}, 022 (1999)
  [arXiv:hep-th/9910053].






\bibitem{Mathur}
  O.~Lunin and S.~D.~Mathur,
   ``Statistical interpretation of Bekenstein entropy for systems with a
  stretched horizon,''
  Phys.\ Rev.\ Lett.\  {\bf 88}, 211303 (2002)
  [arXiv:hep-th/0202072].

  O.~Lunin, S.~D.~Mathur and A.~Saxena,
  ``What is the gravity dual of a chiral primary?,''
  Nucl.\ Phys.\ B {\bf 655}, 185 (2003)
  [arXiv:hep-th/0211292].


  O.~Lunin, J.~M.~Maldacena and L.~Maoz,
  ``Gravity solutions for the D1-D5 system with angular momentum,''
  arXiv:hep-th/0212210.


  S.~D.~Mathur,
  ``The fuzzball proposal for black holes: An elementary review,''
  Fortsch.\ Phys.\  {\bf 53}, 793 (2005)
  [arXiv:hep-th/0502050].



\bibitem{DeWolfe:2001pq}
  O.~DeWolfe, D.~Z.~Freedman and H.~Ooguri,
  ``Holography and defect conformal field theories,''
  Phys.\ Rev.\ D {\bf 66}, 025009 (2002)
  [arXiv:hep-th/0111135].

\bibitem{KPY}
 K.~P.~Yogendran,
  ``D-branes in 2D Lorentzian black hole,''
  JHEP {\bf 0501} (2005) 036
  [arXiv:hep-th/0408114].

\bibitem{Nakayama}
Y.~Nakayama, S.~J.~Rey and Y.~Sugawara,
  ``D-brane propagation in two-dimensional black hole geometries,''
  JHEP {\bf 0509} (2005) 020
  [arXiv:hep-th/0507040].


\end{thebibliography}
\end{document}